\documentclass[conference]{IEEEtran}
\IEEEoverridecommandlockouts
\usepackage[nolist]{acronym}
\usepackage{cite}
\usepackage{amsmath,amssymb,amsfonts}
\usepackage{algorithmic}
\usepackage{graphicx}
\usepackage{tabularx,ragged2e}
\usepackage{textcomp}
\usepackage{xcolor}
\usepackage{todonotes}
\def\BibTeX{{\rm B\kern-.05em{\sc i\kern-.025em b}\kern-.08em
    T\kern-.1667em\lower.7ex\hbox{E}\kern-.125emX}}

\begin{acronym}
\acro{IT}{Information Technology}
\acro{OT}{Operational Technology}
\acro{ICS}{Industrial Control System}
\acro{MBSE}{Model Based Systems Engineering}
\acro{MBT}{Model Based Testing}
\acro{SUT}{System Under Test}
\acro{CVSS}{Common Vulnerability Scoring System}
\acro{CVE}{Common Vulnerabilities and Exposure}
\acro{FSM}{Finite State Machine}
\acro{EFSM}{Extended Finite State Machine}
\acro{TA}{Timed Automata}
\acro{MCUM}{Markov Chain Usage Model}
\acro{V$&$V}{Verification and Validation}
\acro{EDDL}{Electronic Device Description Language}
\acro{SIS}{Safety Instrumented System}
\end{acronym}    
\begin{document}
\title{\LARGE A Model Based Framework for Testing Safety and Security \\ in Operational Technology  Environments\\
}

\author{\IEEEauthorblockN{Mukund Bhole}
\IEEEauthorblockA{\textit{Institute of Computer Engineering} \\
\textit{TU Wien}\\
Vienna, Austria \\
mukund.bhole@tuwien.ac.at}
\and
\IEEEauthorblockN{Wolfgang Kastner}
\IEEEauthorblockA{\textit{Institute of Computer Engineering} \\
\textit{TU Wien}\\
Vienna, Austria \\
wolfgang.kastner@tuwien.ac.at}
\and
\IEEEauthorblockN{Thilo Sauter}
\IEEEauthorblockA{\textit{Institute of Computer Technology,} 
\textit{TU Wien}\\
\textit{Dep. of Integrated Sensor Systems} \\
\textit{Danube University}
Krems,Austria \\\
thilo.sauter@tuwien.ac.at}

}

\maketitle

\begin{abstract}
Today's industrial control systems consist of tightly coupled components allowing adversaries to exploit security attack surfaces from the information technology side, and, thus, also get access to automation devices residing at the operational technology level to compromise their safety functions. To identify these concerns, we propose a model-based testing approach which we consider a promising way to analyze the safety and security behavior of a system under test providing means to protect its components and to increase the quality and efficiency of the overall system. The structure of the underlying framework is divided into four parts, according to the critical factors in  testing of operational technology environments. As a first step, this paper describes the ingredients of the envisioned framework. A system model allows to overview possible attack surfaces, while the foundations of testing and the recommendation of mitigation strategies will be based on process-specific safety and security standard procedures with the combination of existing vulnerability databases. 

\end{abstract}


\section{Introduction}

Over the years, automation systems have been developed considering mainly safety hazards. As automation technologies are nowadays are closely connected to \ac{IT} systems and former borders to \ac{OT} get blurred, adversaries can start a chain of security attacks from the enterprise level and gain access to compromise the safety of \acp{ICS} \cite{b2}. Thus, convergence with increased security flaws at the \ac{IT} level opens doors for harming safety at the \ac{OT} level. Investigating different methodologies is necessary to evaluate the resilience of the system against attacks on the different levels while ensuring that the system considers interdependencies of safety and security.
The paper addresses the mutual dependency of engineering branches on safety and security from a model-based perspective with \ac{MBSE} and safety-security requirement engineering in mind, as these two branches of engineering are not integrated, yet. Integration of these branches can bring benefit on topics such as automated asset management, system complexity management, risk-cost assessment, multidisciplinary team management, and integrated system safety and security evaluation\cite{b21}. The proposed framework supports a methodology to define requirements and analyze the design implementation of an \ac{ICS} before and after the commissioning of \ac{OT} components in the system. Its purpose is to test whether the OT system meets safety and security requirements according to the underlying standards. It integrates four domains into \ac{MBSE}, such as requirements/capabilities \emph{(asset information)}, behavior \emph{(operations/methods)}, structure/architecture \emph{(system modelling)}, verification and validation \emph{(system model testing)}\cite{b1}. Input data essential for our framework are asset information, communication types between assets, operations/methods types executed by assets, and  policies or safety/security measures on asset. The envisioned framework shall be able to retrieve safety and security requirements and their related measures or mitigation strategies from a database. The latter concern best practices to follow and updates to be deployed \cite{b7}. The paper is structured as follows. Section \ref{concept_definitions} outlines the concepts and ingredients of our model-based testing framework. Section \ref{framework} proposes the testing framework approach in brief and sketches an application for a small use case in Section \ref{usecase}. Section \ref{conclusion} draws some concluding remarks and next steps.

\section{Background}
\label{concept_definitions}
\subsection{Model Based System Engineering (MBSE)}
\label{mbse}
\ac{MBSE} is a formalized application of modeling to support system requirements, design analysis, verification, and validation activities starting from conceptual design and continuing throughout development, and later lifecycle phases \cite{b1}. \ac{MBSE} can be beneficial in terms of reduced development cost, system quality, process, and timeline management\cite{b4}.
\subsection{\ac{MBT}}
\label{mbt}
\ac{MBT} is a part of the \ac{MBSE} lifecycle, which works on a deterministic system and demonstrates the implementation's behavior. The most challenging parts for \ac{MBT} are the development of automated test case generation, creation of test data, and definitions of procedures to test the system. \ac{MBT} has proven to increase the quality and efficiency of the system by behavioral analysis of \ac{SUT} models. Once these models have been ensured to reflect the system requirements, these scenarios can serve as ideal test cases in the testing phase \cite{b13}.
\subsection{Vulnerability Assessment}
\label{assessment}
The vulnerability assessment focuses on criticality, OT components' current vulnerabilities, and mitigation. The assessment includes details such as the severity of a threat and a corresponding risk level, determined by, e.g., \ac{CVSS} or \ac{CVE} references, which can be retrieved from vulnerability databases. Moreover, suggestions of mitigation strategies can be part of it, such as available updates and information about patches like name, release date, and update type. Before updates are done, integrity checks need to be  carried out \cite{b7}. 
\subsection{Test Case Generation}
\label{testcase}
The proposed framework will address automated test case generation using state-based models, which may include \acp{FSM}, \acp{EFSM}, UML State Machine Diagrams, \acp{TA}, and \acp{MCUM}. At the moment, individual benefits of these test generation techniques are investigated. \acp{FSM} can be used to generate tests under sandboxing for \ac{SUT} and check if corresponding tests hold for the system and protocol implementation. \acp{EFSM} can generate tests in control and data parts of the system specification. UML can be used to generate a test for multiple processes executing simultaneously in the \ac{SUT} and is also suitable for unit tests based on object-oriented components. \ac{TA} are the best option for test generation in a real-time system with timing constraints and model checking tools. \ac{MCUM} test generation is based on the statistics of execution of states which can be used in complex system components in the \ac{SUT} \cite{b13}.
\subsection{Test Verification \& Validation}
\label{validation}
The test verification and validation (V\&V) is a well-defined approach that evaluates the system throughout its life cycle to the end of product life\cite{b22}. In V\&V of \ac{OT} systems, we considered the most relevant safety and security standards summarized in Table~\ref{tab2}. These standards serve as a basis for the definition of a protection catalog. Based on the catalog, a rule-based system will be developed to deal with the imprecision, modeling method of human behavior, and achieving control of \acp{ICS} by executing a sequence of commands which can or cannot be modeled rigorously\cite{b19}. 
\vspace{-0.2cm}
\begin{table}[!h]
\caption{Safety \& Security Standards}
\resizebox{\columnwidth}{!}{%
\begin{tabular}{|p{1cm}|p{8cm}|}
\hline
\textbf{Criteria}& \textbf{Standards} \\
\hline
Safety  & \textbf{IEC 61511} (Functional safety - Safety instrumented systems for the process industry sector), \textbf{IEC 61311-6} (Programmable controllers - Part 6: Functional safety), \textbf{IEC 61784-3} (Industrial communication networks - Profiles - Part 3: Functional safety fieldbuses - General rules and profile definitions), \textbf{IEC 61508} (methods on how to apply, design, deploy and maintain automatic protection systems called safety-related systems), \textbf{IEC 62061} (Safety of machinery: Functional safety of electrical, electronic and programmable electronic control systems), \textbf{IEC 13849-1} (safety-related parts of a control system)\\
\hline
Safety \& Security  & 
 \textbf{IEC TR 63074} (Safety of machinery - Security aspects related to functional safety of safety-related control systems), \textbf{IEC TR 63069} (Industrial-process measurement, control and automation - Framework for functional safety and security)\\
\hline
Security & \textbf{IEC 62443} (cybersecurity for operational technology in automation and control systems), \textbf{ISO/TR 22100-4} (Guidance to machinery manufacturers for consideration of related IT-security (cyber security) aspects), \textbf{ISO/IEC 27002} (Information technology — Security techniques — Code of practice for information security controls)   \\
\hline
\end{tabular}
}
\label{tab2}
\end{table}

\section{Proposed Framework}
\label{framework}
\begin{figure}[!h]
    \includegraphics[width=8.5cm,height=7.5cm]{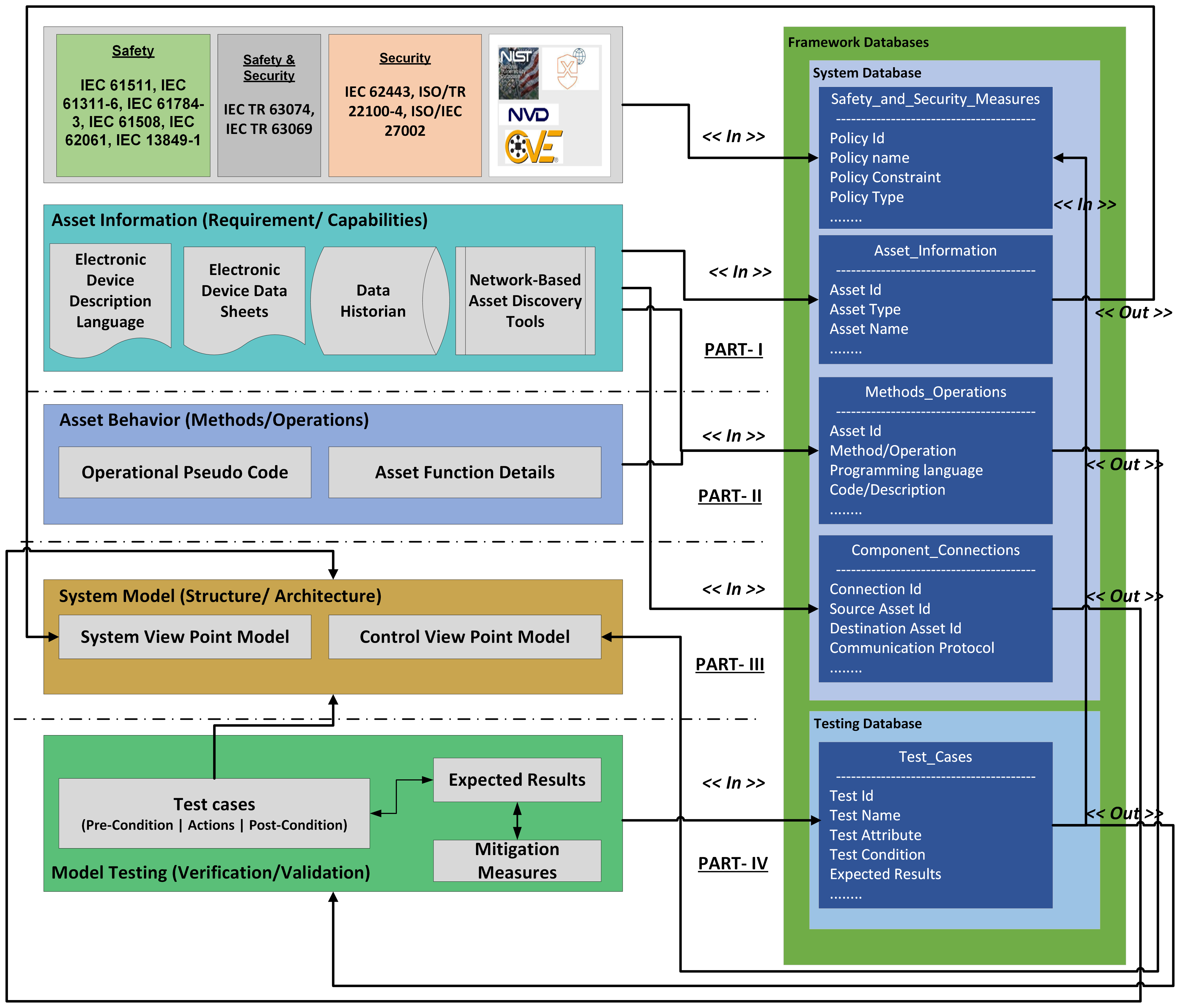}
    \caption{Constituent Parts of the Testing Framework}
    \label{framework_flow}
    \end{figure}

The proposed semi-automated approach to the model-based testing framework for OT environments is divided into four parts illustrated in Fig.\ref{framework_flow}.

\textbf{Part-I Asset Information (Requirements/Capabilities):}
    \label{discovery_approach}
   The asset information can be extracted with a semi-automated approach from different sources such as data historians, electronic data sheets, \ac{EDDL} artifacts, and network-based asset discovery tools. Data extracted from these sources are sketched in Table~\ref{tab4}.
    This information is stored in the system model schema in the tables \textit{asset\_information, methods\_operations}, and  \textit{component\_connections}.

\textbf{Part-II Asset Behavior (Operations/Methods):}
    In this part, operations or functions performed by the asset can be specified (if applicable) in the form of pseudo-code in any supported language for analyzing the asset behavior. Any attribute change from the asset components will impact the methods carrying those attributes, which must be tested. Moreover, it will be possible for developers to check whether the programming language standards are followed, for example, if IEC 61131-3 programming guidelines are met. Additionally, a user can describe the pseudo-code for the operation to be performed by the PLC. The pseudo-code can be validated against the user-defined policies based on programming language standards, such as found in the table \textit{safety\_security\_measures} from the database. These operations/functions are stored in the table \textit{methods\_operations}.
    
\textbf{Part III System Modelling (Structure/Architecture):}
    After collecting details of system components (i.e., asset information, methods/operations, and components connection), the next step is to build a system model from two viewpoints: (1) the System View Point is associated with the management of the sub-system along with the component attributes and holds all static information. (2) the control view point provides the perspective when the system operates and is administered. Thus, the control viewpoint includes the behavioral part of a system generated using table \textit{methods\_operations}, while the system viewpoint holds information about the individual asset components using \textit{asset\_information}. The cardinality and relation between  asset components class is maintained by the table \textit{component\_connection}. The final model will be based on SysML or Automation ML\cite{b20}. 
    
\textbf{Part-IV Model Testing (Verification and Validation):}
    The verification and validation of the system model are performed on the rule-based system description (cf. Subsection \ref{validation}). The rule-based approach allows a tester to check whether the expected ideal condition is met, as defined in the table \textit{test\_cases}.
    These ideal conditions are validated against the test cases to check whether the system satisfies the required pre-condition, actions, and post-condition with the expected results. Test cases are input to the compiled model (cf. Subsection  \ref{testcase}). After testing, if validation fails, required mitigation measures are suggested for successful validation, and changes can be visualized in the system model.
\vspace{-0.3cm}
\begin{table}[!h]
        \caption{Asset Information}
        \vspace{-0.2cm}
        \resizebox{\columnwidth}{!}{%
        \begin{tabular}{|p{2.0cm}|p{6.5cm}|}
        \hline
        \textbf{Approach}& \textbf{Information} \\
        \hline
        Data Historian \cite{b14,b15}  & Plant/process/asset information, asset location, asset specification, sensor readings, product (quality) information, recorded alarms, aggregated data, \dots \\
        \hline
        Electronic Datasheet \cite{b16} & Configuration details, network ports, communication object details, services    \\
        \hline
        EDDL\cite{b17} & Parameter definitions, vendor-specific definitions (manufacturer, device type, revision), parameters, device features    \\
        \hline
        Network Based Asset Discovery Tools\cite{b18} & Asset Information, Connection type, Communication Protocols   \\
        \hline
        \end{tabular}
        }
        \label{tab4}
    \end{table}
\section{Illustrative Example}
\label{usecase}

\begin{figure}[!h]
    \centerline{\includegraphics[width=8.0cm,
    height=6.0cm]{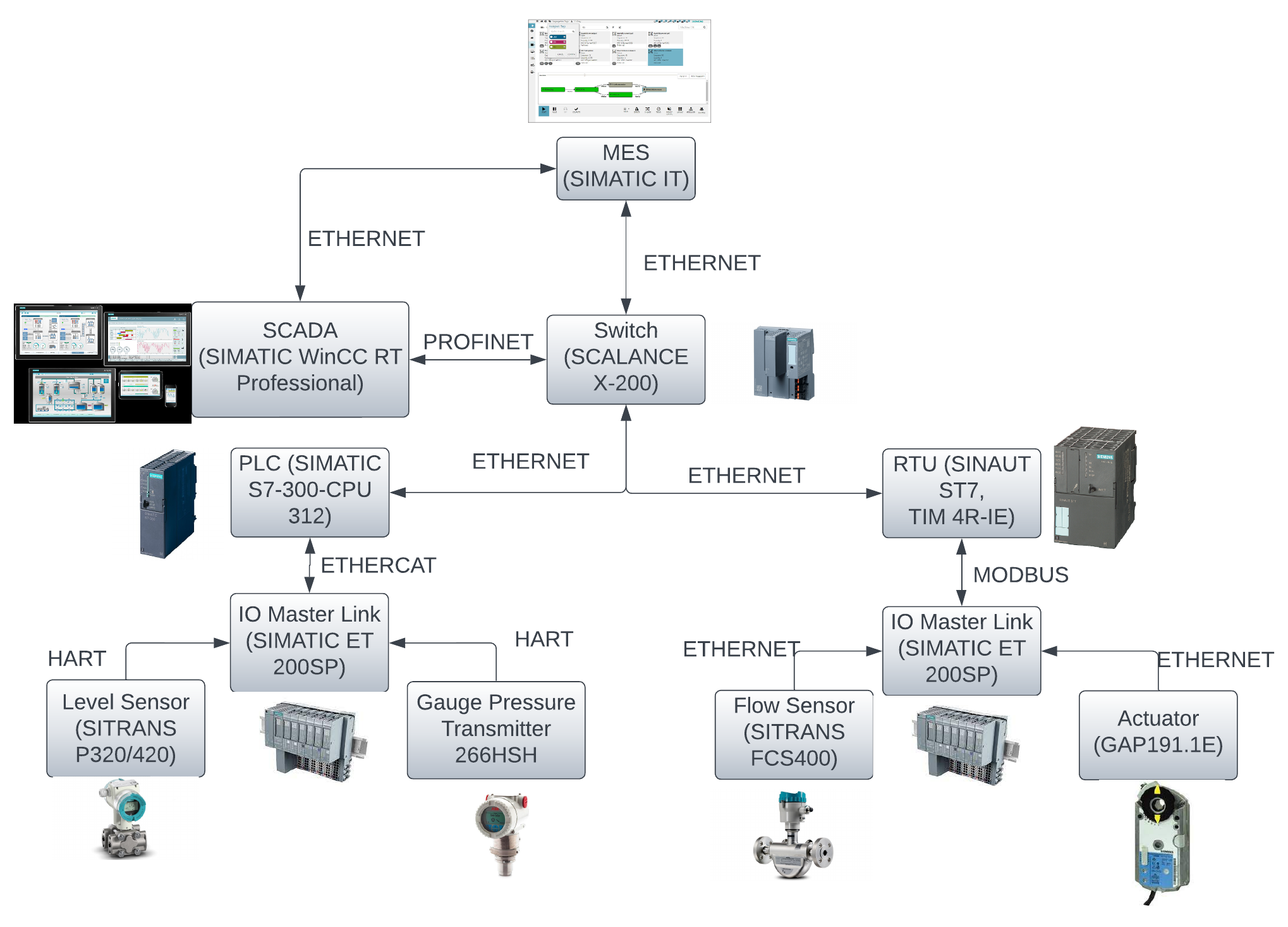}}
    \caption{Illustrative Example}
    \label{example}
    \end{figure}

A simple use case of an \ac{ICS} shall illustrate the implementation of the envisioned framework. The use case (Fig.\ref{example}) consists of OT components up to level 4 of the Purdue model\cite{b23} with sensors, actuators, IO masters, PLCs, RTUs, switches, and workstations. The software side executes firmware, PLC code, SCADA, and MES. Using different techniques mentioned in Table \ref{tab4}, we extract asset information for Part-I and Part-II of the framework. In Part-I, discovered asset type and asset name information shown in Table \ref{ai} are gathered using data historian and network-based asset discovery tools. For the connections of asset components, we can use network-based discovery tools, electronic data sheets, and information from \ac{EDDL} consisting of connection information of the source and destination asset, and communication protocols shown in Table \ref{cc}. In Part II, we extract relevant method information of assets using data historian, electronic data sheets, and \ac{EDDL} data consisting of an operation name. The programming language used to run the operation (if applicable) and a description can be a pseudo code or range of values shown in Table \ref{mo}. In Part III, the framework retrieves extracted system information for generating the system and control viewpoint models in SysML or Automation ML. The system viewpoint model is generated using Tables \ref{ai} and \ref{cc}, while the control viewpoint model is generated using Tables \ref{ai} and \ref{mo}. In Part IV, first, we generate a test case, for example, to test a \ac{SIS} functionality with a PLC (H07) as target. Here test attributes will be an input to automated test case generation (cf. Subsection \ref{testcase}). As shown in Table \ref{tc}, the pre-condition, action, post-condition, and expected result of the test will be fetched. Post-conditions and tester-defined expected results are compared to validate whether the test succeeded or failed. In this example, the expected results and post-conditions are the same, implying a successful validation. In case of a failure, mitigation policies/measures from Table \ref{ssm} are required. As we are testing the \ac{SIS} functionality of the PLC, policies related to safety are recommended (i.e., P01). If the recommended mitigations are fulfilled, then changes are reflected in the model, as shown in Part III of the framework. 

  \begin{table}[!h]\small
        \caption{Safety\_Security\_Measures}
        \resizebox{\columnwidth}{!}{%
        \begin{tabular}{|p{0.6cm}|p{1.5cm}|p{5.0cm}|p{0.8cm}|}
        \hline
        \textbf{Policy ID} & \textbf{Policy Name} & \textbf{Policy Constraint}& \textbf{Policy Type} \\
        \hline
        P01  & IEC 61511: Functional Safety Management & Bypass Power functions; start-up power override; manual shutdown systems; proof-test intervals are documented in the maintenance procedures; diagnostic alarm functions perform as required; utilities are restored, the safety instrumented system returns to the desired state & Safety  \\
        \hline
       P02  & IEC 62443: Patch Management & SIMATIC IT Production Suite (CVE-2018-13804) (Versions V7.1 $<$ V7.1 Upd3) (CVSS Score 9.3); mitigation - restrict network access to affected installations & Security \\
        \hline
        P03  & IEC TR 63074: Security aspects of functional safety & Identify devices covered for vulnerability assessment that could be exploited by threats and influence safety-related control systems   & Safety \& \mbox{Security}  \\
        \hline
        \end{tabular}
        }
        \label{ssm}
    \end{table}
 \begin{table}[!h]\small
        \caption{Test\_Case}
        \resizebox{\columnwidth}{!}{%
        \begin{tabular}{|p{0.4cm}|p{0.7cm}|p{1.2cm}|p{5.5cm}|p{1.3cm}|}
        \hline
        \textbf{Test ID} & \textbf{Test Name} & \textbf{Test Attribute}& \textbf{Test Condition}& \textbf{Expected Results} \\
        \hline
        T01  & SIS Power Safety & PLC (H07) & \textbf{PRE}- Setting and adjustments of PLC logic.
        \textbf{ACTION}- Perform operations - reset, shutdown in different inputs.
        \textbf{POST}- Execute power command and yield output values & Output value is within specified range.\\
        \hline
      T02  & MES Version Test & SIMATIC IT (MES) - S02 &  \textbf{PRE}- – Sandboxing of MES
        \textbf{ACTION}- Current Version Check
        \textbf{POST}- Current Version V7.0 $<$  Updated Version V7.1 & Current Version = Updated Version\\
        \hline
        \end{tabular}
        }
        \label{tc}
        \end{table}

  \begin{table}[!h]
        \caption{Asset\_Information}
        \resizebox{\columnwidth}{!}{%
        \begin{tabular}{|c|c|c|c|}
        \hline
        \textbf{Asset ID} & \textbf{Asset Type} & \textbf{Asset Name} \\
        \hline
        H01  & Hardware & Level Sensor (SITRAN SP320/420)   \\
        \hline
        H02  & Hardware & Gauge Pressure Transmitter 266HSH  \\
        \hline
        H03  & Hardware & Flow Sensor (SITRANS FCS400) \\
        \hline
        H04  & Hardware & Actuator (GAP191.1E) \\
        \hline
        H05  & Hardware & IO Link Master (SIMATIC ET200SP)  \\
        \hline
        H06  & Hardware & IO Link Master (SIMATIC ET200SP)  \\
        \hline
        H07  & Hardware & PLC (SIMATICS7-300-CPU312)  \\
        \hline
        H08  & Hardware & RTU (SINAUTST7,TIM 4R-IE))  \\
        \hline
        H09  & Hardware & Switch (SCALANCE X-200)  \\
        \hline
        H10  & Hardware & Workstation \\
        \hline
        H11  & Hardware & Workstation  \\
        \hline
        S01  & Software & SCADA (SIMATIC WinCC RT Professional)  \\
        \hline
        S02  & Software & MES (SIMATIC IT)\\
        \hline
        S03  & Software & Linux Operating System \\
        \hline
        S04  & Software & Windows 10 Operating System  \\
        \hline
        \end{tabular}
        }
        \label{ai}
    \end{table}
  \begin{table}[!h]
        \caption{Methods\_Operations}
        \begin{tabular}{|p{1.2cm}|p{2.0cm}|p{4.3cm}|}
        \hline
        \textbf{Asset ID} & \textbf{Method/Operation}& \textbf{Code/Description} \\
        \hline
        H02  & Pressure Range   & from 20 mbar to 700 bar  \\
        \hline
        H02  & Temperature Range  & from -40 °C to +100 °C \\
        \hline
        S01  & Monitor Client  & Displays statistics of operations quantifying measurements \& debugging activities \\
        \hline
        S02  & Data Archive   & Achieving values of objects from communication, calculated values, manually entered, or other. \\
        \hline
        \end{tabular}
        
        \label{mo}
    \end{table}
  \begin{table}[!h]
        \caption{Components\_Connections}
        \resizebox{\columnwidth}{!}{%
        \begin{tabular}{|c|c|c|c|}
        \hline
        \textbf{Connection ID} & \textbf{Source Asset ID} & \textbf{Destination Asset ID}& \textbf{Communication Protocol} \\
        \hline
        C01  & H01 & H05 & HART  \\
        \hline
        C02  & H02 & H05 & HART  \\
        \hline
        C03  & H03 & H06 & ETHERNET  \\
        \hline
        C04  & H04 & H06 & ETHERNET  \\
        \hline
        C05  & H05 & H07 & ETHERCAT  \\
        \hline
        C06  & H06 & H08 & MODBUS  \\
        \hline
        C07  & H07 & H09 & ETHERNET  \\
        \hline
        C08  & H08 & H09 & ETHERNET  \\
        \hline
        C09  & H09 & S01+S02+H10 & PROFINET  \\
        \hline
        C10  & H09 & S04+S02+H11 & ETHERNET  \\
        \hline
        C11  & S01+S02+H10 & S04+S02+H11 & ETHERNET  \\
        \hline
        \end{tabular}
        }
        \label{cc}
    \end{table}

\section{Conclusion and Future Work}
\label{conclusion}
The paper attempts to present a model-based testing framework for the safety and security of \ac{OT} systems. The test framework is expected to resolve the safety and security flaws in \acp{ICS} generated due to a lack of synchronization among different development teams and provide the mitigation for flaws detected based on the system model. Fulfilling the prospects of legacy and modern systems on a component level and helping to optimize the system aligned with Industry 4.0.
Further implementation of the framework from design to an actual prototype will be considered important as it will involve fundamental challenges such as dynamic automation technology needs, resource optimization, modeling techniques, and adhering to safety and security standards. We plan to expand the scope of the safety and security measures/policies on a generic level. This way, we might achieve its deployment in multi-domain industries considering the specific needs of those domains, leading to a better evaluation of system components. For this step, the involvement of industry partners is necessary with the ultimate goal of developing a generic meta-model from which a system model in the framework can be derived.
\section*{Acknowledgement}
This paper was supported by TÜV AUSTRIA \#SafeSecLab Research Lab for Safety and Security in Industry, a research cooperation between TU Wien and TÜV AUSTRIA.

\end{document}